\documentclass[final]{amsart}
\usepackage{pdfsync}
\usepackage{amsmath,amssymb}
\usepackage{enumerate}
\usepackage{xspace}
\usepackage{boxedminipage}
\usepackage{algorithmicx}
\usepackage[ruled]{algorithm}
\usepackage{algpseudocode}
\usepackage{listings}
\usepackage{array}
\usepackage{tabularx}
\usepackage{subfigure}
\usepackage{caption}
\usepackage{fullpage}
\usepackage{tristan}

\renewcommand{\vec}[1]{\geovec{#1}}
\renewcommand{\mat}[1]{\geomat{#1}}

\renewcommand{\enorm}[2]{\ensuremath{\Norm{#1}}_{dG,{#2}}}

\renewcommand{\leb}[1]{\ensuremath{\LL^{#1}}}

\numberwithin{equation}{section}

\setboolean{showtodo}{true}
\setboolean{shownotes}{true}
\setboolean{showchanges}{false}
\setboolean{usemathrsfs}{false}
\usepackage{tkz-graph}
\usetikzlibrary{arrows}

\author{
  Georgios Papamikos
}
\address{
  Georgios Papamikos
  \thanks{
School of Mathematics, University of Leeds, Leeds LS2 9JT, UK
{\tt{G.Papamikos@leeds.ac.uk}}.
}}

\author{
 Tristan Pryer
}
\address{
  Tristan Pryer
  \thanks{
    Department of Mathematics and Statistics, Whiteknights, University of Reading, Reading RG6 6AX, UK
    {\tt{T.Pryer@reading.ac.uk}}.
}}

\thanks{{Both authors would like to thank N.Kallinikos and N.Katzourakis for useful discussions. Additionally the constructive comments of both referees are greatly appreciated.} T.P. was partially supported through the EPSRC grant EP/P000835/1. G.P. was supported through a Rado fellowship {and partially through the EPSRC grant EP/P012655/1}. All this support is gratefully acknowledged.}
\title[Lie Symmetry analysis of the 2D $\infty$-Polylaplacian]
      {A Lie symmetry analysis and explicit solutions of the two dimensional $\infty$-Polylaplacian}
      \date{\today}

\begin{document}
\maketitle

\renewcommand{\thefootnote}{\fnsymbol{footnote}} 

\begin{abstract}

  In this work we consider the Lie point symmetry analysis of a
  strongly nonlinear partial differential equation of third order, the
  $\infty$-Polylaplacian, in two spatial dimensions. This
  equation is a higher order generalisation of the $\infty$-Laplacian,
  also known as Aronsson's equation, and arises as the analogue of the
  Euler-Lagrange equations of a second order variational principle in
  $\leb{\infty}$. {We obtain its full symmetry group,
  one dimensional Lie sub-algebras and the corresponding
  symmetry reductions to ordinary differential equations.} Finally, we
  use the Lie symmetries to construct new invariant
  $\infty$-Polyharmonic functions.

\end{abstract}

\section{Introduction}
\label{sec:introduction}

In recent years many partial differential equations (PDEs) that appear as Euler-Lagrange equations in $\leb{\infty}$ variational problems have drawn considerable attention, see \cite{BarronEvansJensen:2008, barron1999, evans2005, Katzourakis:2015} and references therein. These equations are strongly nonlinear elliptic PDEs and appear in many  important applications such as modes for travelling waves in suspension bridges \cite{GyulovMorosanu:2010, LazerMcKenna:1990}, the modelling of granular matter \cite{Igbida:2012}, image processing
\cite{ElmoatazToutainTenbrinck:2015} and game theory \cite{BarronEvansJensen:2008}. 

In this work we study the $\infty$-Polylaplacian equation 
\begin{equation}\label{eq:infpolylap-n}
  \Pi^2_\infty  u:=
  \sum_{i,j = 1}^n
  f[u]_{x_i}  f[u]_{x_j}
  u_{x_i x_j}
  = 0,
\end{equation}
where $u=u(x_1,\ldots,x_n)\in\mathbb{R}$ and $f[u]$ is given by
\begin{equation}
  f[u] :=
  \sum_{i,j = 1}^n (u_{x_i x_j})^2,
\end{equation}
from a Lie-algebraic and computational point of view. As usual the lower index denotes partial differentiation with respect to the corresponding variable.
It is surprising that equation \eqref{eq:infpolylap-n} is a \emph{third order} PDE since it is the formal limit of the $p$-Polylaplacian
\begin{equation}
  \sum_{i,j=1}^n \qp{f[u]^{p/2-1} u_{x_i x_j}}_{x_i x_j}
  =
  0
\end{equation}
as $p\to\infty$ which is a \emph{fourth order}
PDE.
Moreover, the $\infty$-Polylaplacian \eqref{eq:infpolylap-n} has
a connection to an equation that can be seen as a higher order
generalisation of the well known Eikonal equation. Indeed, given the structure
of the operator (\ref{eq:infpolylap-n}), it is clear that solutions to the second order Eikonal-type equation
\begin{equation}
  \label{eq:reducedinfa}
  f[u] = c,
\end{equation}
where $c$ is a constant, are also solutions to the $\infty$-Polylaplacian. 

Equation \eqref{eq:infpolylap-n} can be seen as a higher order generalisation of the $\infty$-Laplacian equation \cite{Aronsson:1965}
\begin{equation}
  \label{eq:inflap}
  \Delta_\infty u
  :=
  \sum_{i,j = 1}^n
  u_{x_i}  u_{x_j}
  u_{x_i x_j}
  = 0,
\end{equation}
also known as Aronsson's equation. The symmetries of equation
\eqref{eq:inflap} with $n=2$ were recently studied and exact solutions
were constructed in \cite{FreireFaleiros:2011}, see also
\cite{Ayanbayev:2018}. Aronsson's equation minimises the Dirichlet
energy functional in $L^p$ as $p\rightarrow\infty$, see also
\cite{KatzourakisPryer:2015, Pryer:2015} for a modern review of the
derivation. There are many difficulties typically encountered in these
variational problems and the study of the associated Euler-Lagrange
equations obtained in this way are notoriously challenging
\cite{Katzourakis:2015}.  Usually solutions are non-classical and need
to be made sense of weakly. The correct notion of weak solutions in
this context is that of \emph{viscosity solutions}
\cite{BhattacharyaDiBenedettoManfredi:1989,Jensen:1993}. In the
context of the $\infty$-Polylaplacian the notion of viscosity
solutions is no longer applicable since we do not have access to a
maximum principle for 3rd order PDEs, from which the solution concept
stems. It is also difficult, due to their complicated form, to
construct exact and physically interesting solutions. {In
  \cite{KatzourakisPryer:2017} equations of this type and the
  structure of their solutions were studied using appropriate
  numerical schemes. One of the goals of this paper is to construct
  new closed form solutions complementing these results.}

For equations that appear in $L^{\infty}$ variational problems, while their analytic properties are thoroughly investigated by many authors, the construction and study of exact solutions is not thoroughly treated. 
There are many successful methods of constructing exact solutions for
nonlinear PDEs. More often than not, these methods rely on some
special algebro-geometric or analytic properties of the PDE,
Darboux-B\"acklund transformations, inverse scattering transform,
Painlev\`e property, etc. see \cite{fordy} for a review. These are
connected to the integrable character of the equation. On the other
hand, Lie group theory is general and makes little assumptions on the
form of the PDE and hence it can be applied to strongly nonlinear and
nonevolution equations such equation \eqref{eq:infpolylap-n}. There
are many modern generalisations of Lie's classical approach. Examples
of such generalisations are the \textit{nonclassical symmetries}
\cite{Bluman1969,Clarkson1994} and \textit{approximate symmetries}
\cite{Baikov1989}, to name a few. A detailed exposition of the
classical theory can be found in the books \cite{Bluman2008,
  Hydon2000, Olver:1993, Ovsiannikov1982, Stephani1989} and the review
papers \cite{Oliveri2010,winternitz1993lie} and in references given
therein. See also \cite{Yaglom1988} for a historic account and
\cite{Dimas2006, Dimas2004sym, hereman1997} for the implementation of
these ideas using computer packages of symbolic algebra. Group
theoretic methods in the study of differential equations have been
applied successfully to problems arising from geometry, general theory
of relativity, gas dynamics, hydrodynamics and many more, see
\cite{Ibragimov1993crc}.


In this work we restrict our attention to the case $n=2$. We use
$x$ and $y$ for the independent variables and so
\eqref{eq:infpolylap-n} simplifies to
\begin{equation}\label{eq:infpolylap-2}
  \begin{split}
    \qp{f[u]_x}^2 u_{xx} + 2 f[u]_{x}f[u]_y u_{xy} + \qp{f[u]_y}^2 u_{yy} & = 0
  \end{split}
\end{equation}
where now
\begin{equation}
  f[u] := (u_{xx})^2+2(u_{xy})^2+(u_{yy})^2.
\label{eq:H-2}
\end{equation}
The aim of this paper is the construction of explicit $\infty$-Polyharmonic functions, i.e. solutions of equation \eqref{eq:infpolylap-2}.
Towards this end we obtain the full Lie symmetry group for both equations \eqref{eq:reducedinfa} and \eqref{eq:infpolylap-2} and we {obtain} one dimensional Lie subalgebras which we use to define appropriate canonical variables and reduce our PDEs to ordinary differential equations (ODEs). Studying the reduced ODEs we construct several new interesting invariant solutions. We also propose a conjecture for the symmetry structure of the general $\infty-$Polylaplacian in $n$ dimensions. With this work we aim to promote group theoretic ideas in the study of these strongly nonlinear problems and their exact solutions, i.e. the $\infty-$Polyharmonic functions \cite{KatzourakisPryer:2016} and find potential minimizers for problems arising in the $L^{\infty}$ variational calculus.

The paper is organised as follows: In the following section we study
the reduced $\infty$-Polylaplacian and consider some of its algebraic properties. 
In Section 3 we briefly
introduce some basics of Lie symmetries of differential equations and
we fix the notation. Moreover, we derive the infinitesimal invariance
conditions for both the $\infty-$Polylaplacian and its reduced
version. We solve the determining equations in Section 4 and thus
obtain the Lie algebras of the full symmetry groups of both
equations. We also present some of the algebraic properties of the Lie
algebras and we obtain the corresponding Lie
symmetries. We also present a conjecture about the Lie symmetries of equations \eqref{eq:infpolylap-n} and \eqref{eq:reducedinfa} for arbitrary $n$.  In Section 5 we {present}
{a partial classification of} inequivalent generators under the action of the adjoint representation, i.e. the action of the symmetry group to its Lie algebra for both equations. We perform the corresponding reductions to ODEs and we construct new invariant solutions. We conclude in Section 6 with a summary and a discussion of our results.

\section{The reduced $\infty-$Polylaplacian equation}

In this section we introduce the reduced $\infty-$Polylaplacian
equation over $\mathbb{R}$, describe some of its properties, and discuss some
possible extensions of this work over algebraically closed fields of characteristic zero.

From the form of equation \eqref{eq:infpolylap-2} it follows that the equation $f[u]=c$, where $c$ is constant, defines a submanifold in the space of solutions of equation \eqref{eq:infpolylap-2}. One can observe that the real valued solutions of $f[u]=c$ which are not affine polynomials in $x$ and $y$ can always be rescaled to real solutions of the following equation
\begin{equation}\label{eq:reducedinf}
    f[u]= u_{xx}^2+2u_{xy}^2+u_{yy}^2 = 1
\end{equation}
and vice versa.
We call equation \eqref{eq:reducedinf} the reduced $\infty$-Polylaplacian. 
Alternatively, one can study the equation $f[u]=c$ for any constant $c\in\mathbb{C}$ over $\mathbb{C}$-valued functions. For example, let $\mathcal{A}=\mathbb{C}[x,y]$ be the ring of polynomials in variables $x$ and $y$ with coefficients in $\mathbb{C}$ and with the usual gradation 
$$
\mathcal{A}=\bigoplus_{k=0}^{\infty}\mathcal{A}_k, \quad \mathcal{A}_i\cdot\mathcal{A}_j\subset\mathcal{A}_{i+j}
$$
where $\mathcal{A}_k$ is the homogeneous component of all polynomials in $\mathcal{A}$ of degree $k$. Then, for simplicity, we can search for solutions of $f[u]=c$ in each subspace $\mathcal{A}_k$. We observe that $\mathcal{A}_0\oplus\mathcal{A}_1\subset \ker f$ and that $f:\mathcal{A}_k\rightarrow \mathcal{A}_{2k-4}$ for all $k\ge 2$. Specifically, for $k=2$ we have that $f[\mathcal{A}_2]=\mathcal{A}_0=\mathbb{C}$ and thus it follows that for any $c\in\mathbb{C}$ the equation $f[u]=c$ admits the solution $u=\alpha x^2+\beta xy+\gamma y^2$ if and only if the parameters $(\alpha,\beta,\gamma)\in\mathbb{C}^3$ are elements of the $1-$parametric family of affine varieties $V(4 \alpha^2+2\beta^2+4\gamma^2-c)$. For $k>2$ and since $f[\mathcal{A}_k]\subset \mathcal{A}_{2k-4}$ it follows that necessarily $c=0$ and thus we only have to consider the equation $f[u]=0$. Moreover, the parameter space associated to a solution $u\in\mathcal{A}_k$ has dimension $\dim\mathcal{A}_k=k+1$ while the image has dimension $\dim\mathcal{A}_{2k-4}=2k-3$. It follows that $f$ maps 
$$ 
u=\sum_{i+j=k}\alpha_{i,j}x^iy^j\mapsto f[u]=\sum_{m+n=2k-4}F_{m,n}(\alpha)x^my^n
$$
where $F_{m,n}$ are homogeneous quadratic polynomials of $\alpha_{i,j}$ and thus $u$ will satisfy equation $f[u]=0$ if and only if $\alpha=(\alpha_{k,0},\alpha_{k-1,1},\ldots,\alpha_{0,k})\in\mathbb{C}^{k+1}$ is an element of the variety $V(I)\subset\mathbb{C}^{k+1}$, where $I$ is the ideal generated by all $F_{m,n}$. Effectively, to find solutions of $f[u]=0$ in $\mathcal{A}_k$ one has to solve $2k-3$ quadratic equations for $k+1$ variables (the parameter space) and thus for $k\geq 5$ the system is overdetermined. In principle these equations can be investigated using Gr\"obner basis, see \cite{Cox1992ideals} and references therein, or numerical schemes. To investigate the existence and the form of solutions in $\mathcal{A}_k$ for all $k$ is an open problem. More generally, the problem of characterising and classifying the solutions of $f[u]=c$ over the ring $\mathbb{F}[x,y]$, where $\mathbb{F}$ is an algebraically closed field of characteristic zero,  is of particular mathematical interest and under current investigation by the authors. In this paper we will not pursuit these ideas any further, instead we assume that $u(x,y)\in\mathbb{R}$ and focus only on the Lie symmetries of the equation \eqref{eq:reducedinf}.

\section{Infinitesimal invariance and determining equations}

In this section we derive the
determining equations of the generators of the symmetry group for both
the $\infty$-Polylaplacian (\ref{eq:infpolylap-2}) and it's reduction
(\ref{eq:reducedinf}).

A Lie point symmetry of equation \eqref{eq:infpolylap-2} is a flow 
\begin{equation}
(\widetilde{x},\widetilde{y},\widetilde{u})=(e^{\epsilon X}x,e^{\epsilon X}y,e^{\epsilon X}u)
\label{eq:flow}
\end{equation}
generated by a vector field
\begin{equation}\label{eq:vf}
X=\xi_1(x,t,u)\frac{\partial}{\partial x}+\xi_2(x,t,u)\frac{\partial}{\partial y}+\eta(x,t,u)\frac{\partial}{\partial u},
\end{equation}
such that $\widetilde{u}(\widetilde{x},\widetilde{y})$ is a solution of \eqref{eq:infpolylap-2} whenever $u(x,y)$ is a solution of \eqref{eq:infpolylap-2}. As usual, we denote by $e^{\epsilon X}$ the \emph{Lie series} $\sum_{k=0}^{\infty}\frac{\epsilon^k}{k!}X^k$ with $X^{k}=XX^{k-1}$ and $X^0=1$.

To find the symmetries of equation \eqref{eq:infpolylap-2} (resp. equation \eqref{eq:reducedinf}) we have to solve the infinitesimal invariance condition for the vector field \eqref{eq:vf}. In the case of equation \eqref{eq:infpolylap-2} (resp. \eqref{eq:reducedinf}) we have to use the third prolongation of $X$ \cite{Bluman2008, Olver:1993, Stephani1989}, namely $X^{(3)}$ (resp. the second  prolonged vector field $X^{(2)}$). The calculations become extremely cumbersome and this is why we use a Mathematica based algebraic package called SYM \cite{Dimas2006, Dimas2004sym} in order to obtain the infinitesimal symmetry conditions. The original infinitesimal symmetry condition for equation  \eqref{eq:infpolylap-2} reads 
\begin{equation}
  \label{infinv}
  X^{(3)}\Pi_{\infty}^2u=0 \quad \text{mod}~(\Pi_{\infty}^2u=0)
\end{equation}
and decomposes to a large overdetermined system of linear PDEs for $\xi_1$, $\xi_2$ and $\eta$ known as \textit{determining equations}. Using computer algebra and algorithms from computational differential algebra \cite{Reid1990, Schwarz2007},  we can prove that the infinitesimal invariance condition (\ref{infinv}) is equivalent  to the following system of 16 equations:
  \begin{eqnarray}
    &\xi_{iu}  =  \xi_{ixx}  =  \xi_{ixy}  =  \xi_{iyy} =  0, \quad i=1,2 \label{eq:deteq1a}
    \\
    &\eta_{xx} = \eta_{xy}  = \eta_{yy} = \eta_{xu} = \eta_{yu} = \eta_{uu} = 0, \label{eq:deteq2a}
    \\ \label{eq:deteq3a}
    &\xi_{1y}+\xi_{2x} = \xi_{1x}-\xi_{2y}  = 0.  
\end{eqnarray}
Similarly, the infinitesimal invariance condition for equation \eqref{eq:reducedinf} 
\begin{equation}
X^{(2)}\left(f[u]-1\right)=0 \quad \text{mod}~(f[u]=1)
\label{eq:infinvb}
\end{equation}
is equivalent to the following system of 12 equations:
  \begin{eqnarray}
    &\xi_{2xx}  =  \xi_{2u}  =  \xi_{1y}+\xi_{2x}  =  \xi_{1x}-\xi_{2y}    =  0, \label{eq:deteq1b}
    \\
    &\eta_{xx}  = \eta_{xy}  = \eta_{yy} = \eta_u-2\xi_{1x} = 0, \label{eq:deteq2b}
    \\ \label{eq:deteq3b}
    & \xi_{1xx} = \xi_{1xy} = \xi_{1xu} = \xi_{1u} = 0.  
\end{eqnarray}   
Solutions of the overdetermined system of linear PDEs \eqref{eq:deteq1a}-\eqref{eq:deteq3a} (resp. \eqref{eq:deteq1b}-\eqref{eq:deteq3b}) will result to the algebra of the symmetry generators \eqref{eq:vf} of equation \eqref{eq:infpolylap-2} (reps. \eqref{eq:reducedinf}).

\section{Lie symmetries of the $\infty$-Polylaplacian}
\label{sec:4}

In this section we focus our attention to the systems of equations
\eqref{eq:deteq1a}-\eqref{eq:deteq3a} and
\eqref{eq:deteq1b}-\eqref{eq:deteq3b}. These equations form an
overdetermined system of linear partial differential equations and
thus it is possible that they only admit the trivial solution
$\xi_1=\xi_2=\eta=0$. This implies that the only {Lie}
symmetry of equation \eqref{eq:infpolylap-2} is the identity
transformation. In what follows we will see that this is not the
case. In this way we obtain the Lie algebra for the symmetry
generators for both equations \eqref{eq:infpolylap-2},
\eqref{eq:reducedinf} and thus, using the Lie series, derive the full
groups of Lie point symmetries for both equations. {At the end of this section we discuss about the discrete symmetries of the equations \eqref{eq:infpolylap-2},
and \eqref{eq:reducedinf}.}

The general solution of the determining equations \eqref{eq:deteq1a}-\eqref{eq:deteq3a} is given by
\begin{equation}\label{eq:sol-gena}
\xi_1=c_1x+c_2y+c_3, \quad \xi_2=-c_2x+c_1y+c_4, \quad \eta=c_5x+c_6y+c_7u+c_8,
\end{equation}
where $c_i$, $i=1,\dots, 8$ are arbitrary real constants.
It follows that the solution \eqref{eq:sol-gena} defines an eight dimensional Lie algebra of generators where the obvious basis is formed by the following vector fields
\begin{eqnarray}
& X_1=\frac{\partial }{\partial x},~ X_2=\frac{\partial}{\partial y},~ X_3=-y\frac{\partial}{\partial x}+x\frac{\partial}{\partial y}, X_4=x\frac{\partial}{\partial x}+y\frac{\partial}{\partial y}, \label{eq:8basisa1}\\
& X_5=u\frac{\partial}{\partial u},~X_6=x\frac{\partial}{\partial u},~X_7=y\frac{\partial}{\partial u},~X_8=\frac{\partial }{\partial u}.\label{eq:8basisa2}
\end{eqnarray}
Similarly, for equation \eqref{eq:reducedinf}, we have that the general solution of the determining equations \eqref{eq:deteq1b}-\eqref{eq:deteq3b} is given by
\begin{equation}\label{eq:sol-genb}
\xi_1=c_1x-c_2y+c_3, \quad \xi_2=c_2x+c_1y+c_4, \quad \eta=c_5x+c_6y+2c_1u+c_7,
\end{equation}
where $c_i$, $i=1,\dots, 7$ are arbitrary real constants.
The Lie algebra of vector fields defined by the solution \eqref{eq:sol-genb} is similar to the algebra spanned by the vector fields \eqref{eq:8basisa1}-\eqref{eq:8basisa2}. It is a  seven dimensional Lie algebra spanned by the vector fields
\begin{eqnarray}
& Y_1=\frac{\partial }{\partial x},~ Y_2=\frac{\partial}{\partial y},~ Y_3=-y\frac{\partial}{\partial x}+x\frac{\partial}{\partial y}, Y_4=x\frac{\partial}{\partial x}+y\frac{\partial}{\partial y}+2u\frac{\partial}{\partial u}, \label{eq:7basisb1}\\
& Y_5=x\frac{\partial}{\partial u},~Y_6=y\frac{\partial}{\partial u},~Y_7=\frac{\partial }{\partial u}.\label{eq:7basisb2}
\end{eqnarray}
The reason for this symmetry breaking is because the reduced equation
(\ref{eq:reducedinf}) is not homogeneous and hence the equation only
admits the scaling that makes each individual term, $u_{xx}$,
$u_{xy}$ and $u_{yy}$ invariant, i.e. the symmetry generated by $Y_4$.

We denote the eight dimensional real Lie algebra by $\mathfrak{g}$ and the seven dimensional
real Lie algebra by $\mathfrak{h}$, viz.
\begin{equation}
\mathfrak{g}=\text{Span}\lbrace X_i,~i=1,\ldots, 8\rbrace, \quad \mathfrak{h}=\text{Span}\lbrace Y_i,~i=1,\ldots, 7\rbrace.
\label{eq:lie-alg}
\end{equation}   
Then it follows that equation \eqref{eq:infpolylap-2} admits the symmetry group generated by $\mathfrak{g}$ while the symmetries of equation \eqref{eq:reducedinf} are generated by $\mathfrak{h}$.  Moreover, since for both equations $\xi_{1u}=\xi_{2u}=0$, it follows that the symmetry transformations of both \eqref{eq:infpolylap-2} and \eqref{eq:reducedinf} are \textit{fibre preserving transformations}.

Both Lie algebras $\mathfrak{g}$ and $\mathfrak{h}$ are \textit{solvable}. Indeed, we have that for both algebras the \textit{derived series}
$$
\mathfrak{g}^{(n)}=\left[\mathfrak{g}^{(n-1)},\mathfrak{g}^{(n-1)}\right], \quad \mathfrak{g}^{(0)}=\mathfrak{g}
$$ 
terminate to the trivial Lie algebra $\mathfrak{0}=\lbrace 0\rbrace$ for a positive integer $n$. As usual $[\cdot,\cdot]$ denotes the commutator of vector fields which is the \textit{Lie bracket} of $\mathfrak{g}$ and $\mathfrak{h}$. The first derived algebra, which is an ideal of $\mathfrak{g}$, is
$$
\mathfrak{g}^{(1)}=\left[\mathfrak{g},\mathfrak{g}\right]=\text{Span}\lbrace X_1,X_2,X_6,X_7,X_8 \rbrace
$$
and 
$$
\mathfrak{h}^{(1)}=\text{Span}\lbrace Y_1,Y_2,Y_5,Y_6,Y_7\rbrace
$$
as can be verified by inspecting Table \ref{tab:Lie-table1} and Table \ref{tab:Lie-table2}.
\begin{center}
  \begin{tabular}{ | c | c  c  c  c  c  c  c c |}   
    \hline
    $\left[X_i,X_j\right]$ & $X_1$ & $X_2$ & $X_3$ & $X_4$ & $X_5$ & $X_6$ & $X_7$ & $X_8$ \\ \hline
    $X_1$ & 0 & 0 & $X_2$ & $X_1$ & 0 & $X_8$ & 0 & 0 \\ 
    $X_2$ & 0 & 0 & $-X_1$ & $X_2$ & 0 & 0 & $X_8$ & 0 \\      
    $X_3$ & $-X_2$ & $X_1$ & 0 & 0 & 0 & $-X_7$ & $X_6$ & 0 \\ 
    $X_4$ & $-X_1$ & $-X_2$ & 0 & 0 & 0 & $X_6$ & $X_7$ & 0 \\ 
    $X_5$ & 0 & 0 & 0 & 0 & 0 & $-X_6$ & $-X_7$ & $-X_8$ \\ 
    $X_6$ & $-X_8$ & 0 & $X_7$ & $-X_6$ & $X_6$ & 0 & 0 & 0 \\ 
    $X_7$ & 0 & $-X_8$ & $-X_6$ & $- X_7$ & $X_7$ & 0 & 0 & 0 \\ 
    $X_8$ & 0 & 0 & 0 & 0 & $X_8$ & 0 & 0 & 0 \\ \hline 
  \end{tabular}
\captionof{table}{  {\label{tab:Lie-table1}}
  Commutation relations of the Lie algebra $\mathfrak{g}$.}
\end{center}
Similarly, we have that 
$$
\mathfrak{g}^{(2)}=\text{Span}\lbrace X_8\rbrace, \quad \mathfrak{h}^{(2)}=\text{Span}\lbrace Y_7\rbrace
$$
and thus $\mathfrak{g}^{(3)}=\mathfrak{h}^{(3)}=\mathfrak{0}$.

\begin{center} 
  \begin{tabular}{ | c | c  c  c  c  c  c  c |}   
    \hline
    $\left[Y_i,Y_j\right]$ & $Y_1$ & $Y_2$ & $Y_3$ & $Y_4$ & $Y_5$ & $Y_6$ & $Y_7$ \\ \hline
    $Y_1$ & 0 & 0 & $Y_2$ & $Y_1$ & $Y_7$ & 0 & 0 \\ 
    $Y_2$ & 0 & 0 & $-Y_1$ & $Y_2$ & 0 & $Y_7$ & 0 \\      
    $Y_3$ & $-Y_2$ & $Y_1$ & 0 & 0 & $-Y_6$ & $Y_5$ & 0 \\ 
    $Y_4$ & $-Y_1$ & $-Y_2$ & 0 & 0 & $-Y_5$ & $-Y_6$ & $-2 Y_7$ \\ 
    $Y_5$ & $-Y_7$ & 0 & $Y_6$ & $Y_5$ & 0 & 0 & 0 \\ 
    $Y_6$ & 0 & $-Y_7$ & $-Y_5$ & $Y_6$ & 0 & 0 & 0 \\ 
    $Y_7$ & 0 & 0 & 0 & $2 Y_7$ & 0 & 0 & 0 \\ \hline
  \end{tabular}
  \captionof{table}{
    {    \label{tab:Lie-table2}}
    Commutation relations of the Lie algebra $\mathfrak{h}$.}
\end{center}

{Since the Lie algebra $\mathfrak{g}$ is solvable it admits a unique maximal ideal which is nilpotent and is called \textit{nilradical}. It can be verified that $\mathfrak{g}$ can be written as the semi-direct sum $\mathfrak{g}=\mathfrak{g}_1\oplus_s\mathfrak{g}_2$, where $\mathfrak{g}_1=\text{Span}(X_1,X_2,X_6,X_7,X_8)$ is its nilradical, $\mathfrak{g}_2=\text{Span}(X_3,X_4,X_5)$ is the three dimensional abelian subalgebra, and thus the following relations hold
\begin{equation}
\left[\mathfrak{g}_1,\mathfrak{g}_1\right]\subset \mathfrak{g}_1, \quad \left[\mathfrak{g}_2,\mathfrak{g}_2\right]=\mathfrak{0}, \quad \left[\mathfrak{g}_1,\mathfrak{g}_2\right]\subset\mathfrak{g}_1.
\label{eq:directsumofg}
\end{equation}
The nilradical $\mathfrak{g}_1$ has appeared in a classification of five dimensional nilpotent Lie algebras. Indeed, $\mathfrak{g}_1$ is isomorphic to the Lie algebra $\mathfrak{n}_{5,3}$, see page 231 in \cite{SnoblWinternitz:2014}. Similarly, the Lie algebra $\mathfrak{h}$ can be decomposed as a semi-direct sum of its nilradical $\mathfrak{h}_1=\text{Span}(Y_1,Y_2,Y_5,Y_6,Y_7)$ and of the abelian subalgebra $\mathfrak{h}_2=\text{Span}(Y_3,Y_4)$.}

It is important to mention that the Lie subalgebra $\text{Span}\lbrace X_1,X_2,X_3\rbrace$ is the Euclidean Lie algebra $e(2)$ formed by the \textit{Killing vector fields} of $\mathbb{R}^2$. Thus equation \eqref{eq:infpolylap-2}  inherits the symmetries of the metric structure of $\mathbb{R}^2$ as it was also pointed out for the Aronsson's equation \eqref{eq:inflap} in \cite{FreireFaleiros:2011}. The extension ${\widetilde{\mathfrak{g}}}=\text{Span}\lbrace X_1,X_2,X_3,X_4,X_5\rbrace$ of the Lie algebra $e(2)$ is still a Lie subalgebra of $\mathfrak{g}$. Finally, as can be seen by the Table \ref{tab:Lie-table1}, the Lie subalgebra $\text{Span}\lbrace X_6,X_7,X_8\rbrace$ forms an abelian ideal of $\mathfrak{g}$. 
Similar things can be said for the Lie algebra $\mathfrak{h}$. Here ${\widetilde{\mathfrak{h}}}=\text{Span}\lbrace Y_1,Y_2,Y_3,Y_4 \rbrace$ and $\text{Span}\lbrace Y_5,Y_6,Y_7 \rbrace$ is also abelian ideal. {We will use the subalgebras $\widetilde{\mathfrak{g}}$ and $\widetilde{\mathfrak{h}}$ in the next section in the construction of invariant solutions.} The identification of the structural properties of the symmetry Lie algebras is important since they can be used for deciding whether or not another PDE can be mapped to the equation at hand. {Additionally, depending on the structural properties of the Lie algebra (simple, semi-simple, etc.) there are methods for the complete classification of its subalgebras. For example, it is possible to use the decomposition of $\mathfrak{g}$ \eqref{eq:directsumofg} in order to fully classify all its one-dimensional subalgebras into conjugation classes using algorithms presented in \cite{PateraWinternitzZassenhaus:1975}, see also \cite{Winternitz:1990} for a review.} 

Using the Lie series we find that the full group of Lie point symmetries of 
\eqref{eq:infpolylap-2} is generated by:
\begin{equation*}
\begin{split}  
      G_1:(x,y,u)  \mapsto   (\widetilde{x},\widetilde{y},\widetilde{u})&=(x+\epsilon,y,u),\\
      G_2:(x,y,u)  \mapsto   (\widetilde{x},\widetilde{y},\widetilde{u})&=(x,y+\epsilon,u),\\
      G_3:(x,y,u)  \mapsto   (\widetilde{x},\widetilde{y},\widetilde{u})&=(x \text{ cos}\epsilon-y \text{ sin}\epsilon,x\text{ sin}\epsilon+y\text{ cos}\epsilon,u),\\
      G_4:(x,y,u)  \mapsto   (\widetilde{x},\widetilde{y},\widetilde{u})&=(e^{\epsilon}x,e^{\epsilon}y,u),\\
      G_5:(x,y,u)  \mapsto  (\widetilde{x},\widetilde{y},\widetilde{u})&=(x,y,e^{\epsilon}u),\\
      G_6:(x,y,u)  \mapsto   (\widetilde{x},\widetilde{y},\widetilde{u})&=(x,y,u+\epsilon x),\\
      G_7:(x,y,u)  \mapsto   (\widetilde{x},\widetilde{y},\widetilde{u})&=(x,y,u+\epsilon y),\\
      G_8:(x,y,u)  \mapsto   (\widetilde{x},\widetilde{y},\widetilde{u})&=(x,y,u+\epsilon).
\end{split}
\end{equation*}
Using the Lie symmetries $G_i$ we can construct new solutions from known solutions.  It follows that if $u=g(x,y)$ is a solution of equation \eqref{eq:infpolylap-2} then the following:
$$
\begin{array}{lcl}
u=g(x,y)+\epsilon x, & & u=g(x-\epsilon,y),\\
u=g(x,y)+\epsilon y, & & u=g(x,y-\epsilon),\\
u=g(x,y)+\epsilon, & & u=g(x \text{ cos}\epsilon+y \text{ sin}\epsilon,-x\text{ sin}\epsilon+y\text{ cos}\epsilon),\\
u=e^{\epsilon}g(x,y), & & u=g(e^{-\epsilon}x,e^{-\epsilon}y)
\end{array}
$$
are also solutions for all $\epsilon\in\mathbb{R}$.
We obtain a similar symmetry structure for the reduced
$\infty-$Polylaplacian equation \eqref{eq:reducedinf} with the only
difference being a breaking of the scaling symmetries, resulting in
one fewer. More precisely, $G_1-G_3$ and $G_6-G_8$ are also Lie symmetries of \eqref{eq:reducedinf} but $G_4$ and $G_5$ are not. Instead, equation \eqref{eq:reducedinf} admits the following scaling symmetry
$$
H:(x,y,u)\mapsto(\widetilde{x},\widetilde{y},\widetilde{u})=(e^{\epsilon}x,e^{\epsilon}y,e^{2\epsilon}u)
$$
which implies that if $u=g(x,y)$ is a solution of
\eqref{eq:reducedinf} then so is 
$$
u=e^{2\epsilon}g(e^{-\epsilon}x,e^{-\epsilon}y),
$$ 
for all $\epsilon \in \reals$.
The scaling $H$ can be seen as the composition of the
scaling transformations $G_4$ and $G_5$ making
each differential monomial on the left hand side of equation
\eqref{eq:reducedinf} invariant under the action of $H$.

Equations \eqref{eq:infpolylap-2} and \eqref{eq:reducedinf} also admit
discrete symmetries. In particular, both equations
\eqref{eq:infpolylap-2} and \eqref{eq:reducedinf} are invariant under
the permutation $\sigma$ of the independent variables $x$ and $y$, as
well as the $x$-reflection $\rho:x\mapsto -x$. Obviously, the
$y$-reflection is also a symmetry of both equations and can be
expressed as $\sigma\circ \rho\circ \sigma$. These are also symmetries
of the general $\infty$-Polylaplacian \eqref{eq:infpolylap-n} in any
dimension $n$. Moreover, \eqref{eq:reducedinf}, while not scaling
invariant in the $u$-direction, remains invariant under the reflection
$u\mapsto-u$.

The study of the Lie symmetry structure of the $\infty$-Polylaplacian
for dimensions $n\ge 3$ and its corresponding reduction
(\ref{eq:reducedinf}) is an open problem which is left for future
work. Nevertheless, we formulate the following:

\vspace{0.2cm}

\noindent\textit{Conjecture}: The Lie
algebra of the symmetry generators of $\infty-$Polylaplacian equation
\eqref{eq:infpolylap-n} in $n$ independent variables has dimension
$3+n(n+3)/2$ and it is spanned by the vector fields related to translations {in the independent variables}, scalings and {affine linear translations in the dependent variable}
\begin{equation}
  \frac{\partial}{\partial x_i}, \quad
  \frac{\partial}{\partial u},\quad
  u\frac{\partial}{\partial u},\quad
  \sum_{j=1}^nx_j\frac{\partial}{\partial x_j}, \quad
  x_i\frac{\partial}{\partial u},\quad
  i = 1,\ldots,n,
\end{equation}
as well as rotation symmetries generated by
\begin{equation} 
  -x_i\frac{\partial}{\partial x_j}+x_j\frac{\partial} {\partial x_i},  ~1\leq i < j \le n.
\end{equation}
Similarly, the symmetry algebra of the reduced $\infty-$Polylaplacian
equation (\ref{eq:reducedinfa}) in $n$ independent variables has
dimension $2+n(n+3)/2$ and is spanned by the same generators for
translation, rotations and {affine linear translations in the dependent variable} but with a scaling symmetry generated by
\begin{equation}
  \sum_{j=1}^nx_j\frac{\partial}{\partial
  x_j}+2u\frac{\partial}{\partial u}.
\end{equation}
The above conjecture has been verified by the authors for $n=3$.

\section{Symmetry reductions and invariant solutions}
 
In this section we are concerned with the symmetry reductions and the
construction of group invariant solutions of equations
\eqref{eq:infpolylap-2} and \eqref{eq:reducedinf}. We construct
solutions that are invariant under one dimensional subgroups acting
non-trivially on the independent variables. {More
  specifically, we focus on the symmetry subalgebras
  $\widetilde{\mathfrak{g}}$ and $\widetilde{\mathfrak{h}}$ and we
  classify their one dimensional Lie subalgebras into equivalence
  classes under the action of the corresponding group. As already
  mentioned in section \ref{sec:4} by focussing on these subalgebras
  we will not obtain a full classification, however the problem is
  tractable and we are focussing on symmetries that have a physical
  meaning. In particular, some of the explicit solutions of the
  corresponding reduced ODEs are related to
  the results of numerical experimentation in
  \cite{KatzourakisPryer:2017}. The complete classification is left as a
  future work.}


We first consider the reductions of equation \eqref{eq:infpolylap-2}
using one dimensional subalgebras of $\mathfrak{g}$ spanned by
$X_i,~i=1,\ldots, 5$. To classify all the one dimensional subalgebras
of $\widetilde{\mathfrak{g}}=\text{Span}\lbrace X_1,\ldots, X_5 \rbrace$
we need to consider the action of the adjoint representation of the
symmetry group of equation \eqref{eq:infpolylap-2} on
$\widetilde{\mathfrak{g}}$. The \textit{adjoint representation} of a Lie group to its algebra is a group action and is defined by conjugacy as follows
$$
\text{Ad}_{\exp\epsilon X}(Y)=e^{\epsilon X}Ye^{-\epsilon X}=e^{\epsilon \text{ad}_X(Y)}=Y+\epsilon \text{ad}_X(Y)+\frac{\epsilon^2}{2!}\text{ad}^2_X(Y)+\cdots,
$$
where $X$ and $Y$ are elements of the Lie algebra and $\text{ad}_X(Y)=\left[X,Y\right]$, see for example \cite{Olver:1993}.
For the sake of completeness we present the adjoint representation of the symmetry group of \eqref{eq:infpolylap-2} on its whole Lie algebra $\mathfrak{g}$ in Table \ref{tab:Lie-table3} and of the symmetry group of \eqref{eq:reducedinf} to $\mathfrak{h}$ in Table \ref{tab:Lie-table4}.
\newcolumntype{R}[1]{>{\raggedleft\let\newline\\\arraybackslash\hspace{1pt}}m{#1}}
\setlength\tabcolsep{4pt}
\begin{tiny}
\begin{center} 
    \begin{tabular}{ |c| c  c  c  c  c  c  c c |}   
    \hline
    $\text{Ad}$ & $X_1$ & $X_2$ & $X_3$ & $X_4$ & $X_5$ & $X_6$ & $X_7$ & $X_8$ \\ \hline
    $X_1$ & $X_1$ & $X_2$
    &
    $X_3+\epsilon X_2$ & $X_4+\epsilon X_1$ & $X_5$ & $X_6+\epsilon X_8$ & $X_7$ & $X_8$ \\ 
    $X_2$ & $X_1$ & $X_2$ & $X_3-\epsilon X_1$ & $X_4+\epsilon X_2$ & $X_5$ & $X_6$ & $X_7+\epsilon X_8$ & $X_8$ \\      
    $X_3$ & 
    \begin{tabular}{c}
    $\text{cos}\epsilon X_1$\\
    $-\text{sin}\epsilon X_2$
    \end{tabular}
     & 
     \begin{tabular}{c}
     $\text{sin}\epsilon X_1$\\
     $+\text{cos}\epsilon X_2$
     \end{tabular}
     & $X_3$ & $X_4$ & $X_5$ 
     & 
     \begin{tabular}{c}
     $\text{cos}\epsilon X_6$\\
     $-\text{sin}\epsilon X_7$
     \end{tabular}
     & 
     \begin{tabular}{c}
     $\text{sin}\epsilon X_6$\\
     $+\text{cos}\epsilon X_7$
     \end{tabular}          
     & $X_8$ \\ 
    $X_4$ & $e^{-\epsilon} X_1$ & $e^{-\epsilon} X_2$ & $X_3$ & $X_4$ & $X_5$ & $e^{\epsilon}X_6$ & $e^{\epsilon}X_7$ & $X_8$ \\ 
    $X_5$ & $X_1$ & $X_2$ & $X_3$ & $X_4$ & $X_5$ & $e^{-\epsilon}X_6$ & $e^{-\epsilon}X_7$ & $e^{-\epsilon}X_8$ \\ 
    $X_6$ & $X_1-\epsilon X_8$ & $X_2$ & $X_3+\epsilon X_7$ & $X_4-\epsilon X_6$ & $X_5+\epsilon X_6$ & $X_6$ & $X_7$ & $X_8$ \\ 
    $X_7$ & $X_1$ & $X_2-\epsilon X_8$ & $X_3-\epsilon X_6$ & $X_4-\epsilon X_7$ & $X_5+\epsilon X_7$ & $X_6$ & $X_7$ & $X_8$ \\ 
    $X_8$ & $X_1$ & $X_2$ & $X_3$ & $X_4$ & $X_5+\epsilon X_8$ & $X_6$ & $X_7$ & $X_8$ \\ \hline 
  \end{tabular}
    \captionof{table}{
      \label{tab:Lie-table3}
      The $Ad_{\exp\epsilon X_i}X_j$ is shown in the $(i,j)$ entry of the table.}
\end{center}
\end{tiny}
\begin{tiny}
\begin{center} 
  \begin{tabular}{ | c | c  c  c  c  c  c  c |}   
    \hline
    $\text{Ad}$ & $Y_1$ & $Y_2$ & $Y_3$ & $Y_4$ & $Y_5$ & $Y_6$ & $Y_7$  \\ \hline
    $Y_1$ & $Y_1$ & $Y_2$ & $Y_3+\epsilon Y_2$ & $Y_4+\epsilon Y_1$ & $Y_5+\epsilon Y_7$ & $Y_6$ & $Y_7$  \\ 
    $Y_2$ & $Y_1$ & $Y_2$ & $Y_3-\epsilon Y_1$ & $Y_4+\epsilon Y_2$ & $Y_5$ & $Y_6+\epsilon Y_7$ & $Y_7$  \\      
    $Y_3$ & 
    \begin{tabular}{c}
    $\text{cos}\epsilon Y_1$\\
    $-\text{sin}\epsilon Y_2$
    \end{tabular}
     & 
     \begin{tabular}{c}
     $\text{sin}\epsilon Y_1$\\
     $+\text{cos}\epsilon Y_2$
     \end{tabular}
     & $Y_3$ & $Y_4$  
     & 
     \begin{tabular}{c}
     $\text{cos}\epsilon Y_5$\\
     $-\text{sin}\epsilon Y_6$
     \end{tabular}
     & 
     \begin{tabular}{c}
     $\text{sin}\epsilon Y_5$\\
     $+\text{cos}\epsilon Y_6$
     \end{tabular}          
     & $Y_7$ \\ 
    $Y_4$ & $e^{-\epsilon} Y_1$ & $e^{-\epsilon} Y_2$ & $Y_3$ & $Y_4$ & $e^{-\epsilon}Y_5$ & $e^{-\epsilon}Y_6$ & $e^{-2\epsilon}Y_7$ \\ 
    $Y_5$ & $Y_1-\epsilon Y_7$ & $Y_2$ & $Y_3+\epsilon Y_6$ & $Y_4+\epsilon Y_5$ & $Y_5$ & $Y_6$ & $Y_7$  \\ 
    $Y_6$ & $Y_1$ & $Y_2-\epsilon Y_7$ & $Y_3-\epsilon Y_5$ & $Y_4+\epsilon Y_6$ & $Y_5$ & $Y_6$ & $Y_7$  \\ 
    $Y_7$ & $Y_1$ & $Y_2$ & $Y_3$ & $Y_4+2\epsilon Y_7$ & $Y_5$ & $Y_6$ & $Y_7$ \\ \hline 
\end{tabular}
  \captionof{table}{
    \label{tab:Lie-table4}
    The $Ad_{\exp\epsilon Y_i}Y_j$ is shown in the $(i,j)$ entry of the table.}
\end{center}  
\end{tiny}

Any one dimensional subalgebra of $\widetilde{\mathfrak g}$ is
equivalent, under the adjoint representation, to one of the following cases:
$$
\begin{array}{clcl}
    (A1) & X_1, & (A5) & \alpha X_3+X_5, \\
    (A2) & X_3, & (A6) & \alpha X_4+X_5,\\
    (A3) & X_4, & (A7) & \gamma X_1+\alpha X_3+X_4, \\ 
    (A4) & \gamma X_1+X_5  & (A8) & \gamma X_1+\alpha X_3+\beta X_4+X_5,
\end{array}
$$
where $\gamma\in\lbrace 0,1\rbrace$ and $\alpha, \beta\in\mathbb{R}\backslash\lbrace 0\rbrace$. Starting with a general element of $\widetilde{\mathfrak{g}}$ of the form
$$
X=\alpha_1X_1+\alpha_2X_2+\alpha_3X_3+\alpha_4X_4+\alpha_5X_5
$$ 
we can use all $\text{Ad}_{\exp\epsilon X_i}$ in order to simplify as much as possible and effectively classify all different 
one dimensional subalgebras of $\widetilde{\mathfrak{g}}$. The adjoint
action $\text{Ad}$ induces an action on the coefficients $\alpha_i$,
i.e. on $\mathbb{R}^5$. We observe that $\alpha_3$,
$\alpha_4$ and $\alpha_5$ are invariants of the induced action. This
implies that we can classify all inequivalent vector fields
according to whether these invariants are zero or not. Moreover, we
can rescale the vector field $X$, use the permutation
symmetry $\sigma$ and the reflection symmetries
$\rho$ and $\sigma\circ \rho \circ \sigma$ to identify some subcases and thus
simplify our classification list. For example, in the case where $\alpha_3=\alpha_4=\alpha_5=0$ we act with $Ad_{\exp\epsilon X_3}$ to $X$ and we obtain $\widetilde{X}=(\alpha_1\cos\epsilon+\alpha_2 \sin\epsilon)X_1+(\alpha_2\cos\epsilon-\alpha_1 \sin\epsilon)X_2$. Choosing $\epsilon=\text{arctan}(\alpha_2\alpha_1^{-1})$ and multiplying by a constant factor we obtain $X_1$. The other cases are obtained in a similar manner but the calculations are omitted for simplicity. The interested reader can find more details on such constructions as well as simpler examples in \cite{Hydon2000, Olver:1993, Ovsiannikov1982}.

Similar considerations hold for the symmetry algebra of equation \eqref{eq:reducedinf}
$\widetilde{\mathfrak{h}}=\text{Span}\lbrace Y_1,\ldots, Y_4\rbrace$. In this case any one dimensional subalgebra of $\widetilde{\mathfrak h}$ is
equivalent to one of the following cases:
$$
\begin{array}{clcl}
    (B1) & Y_1, & (B3) & Y_4, \\
    (B2) & Y_3, & (B4) & \gamma Y_1+\alpha Y_3+Y_4,\\
\end{array}
$$
where $\gamma\in\lbrace 0 ,1\rbrace$ and $\alpha\in\mathbb{R}\backslash\lbrace 0\rbrace$.
To prove this we use similar arguments. Beginning with a general element of
$\widetilde{\mathfrak h}$ of the form
$$
Y=\beta_1Y_1+\beta_2Y_2+\beta_3Y_3+\beta_4Y_4
$$ 
we classify all inequivalent cases. In this case the invariants of the induced action are $\beta_3$ and $\beta_4$.

\subsection{Invariant solutions via symmetry reductions}

We proceed by first considering the symmetry reductions to ODEs and
then continue constructing new solutions, of equation \eqref{eq:infpolylap-2}, which are invariant under the symmetry transformations corresponding to the vector fields A1-A8. We do the same for equation \eqref{eq:reducedinf}. We first
consider the reductions and solutions of equation
\eqref{eq:infpolylap-2}.

\subsubsection{$1$}
Solutions of \eqref{eq:infpolylap-2} which are invariant under the
symmetry generated by $X_1$ are of the form $u=g(y)$. This implies
that $g$ is a solution of the trivial ODE $g''(y)^3g'''(y)^2=0$ and
thus it follows that $u=c_1y^2+c_2y+c_3$ satisfies the
$\infty-$Polylaplacian. Since equation \eqref{eq:infpolylap-2} admits
the permutation $\sigma$ and also contains
derivatives of at least second order it is easy to verify that the
general quadratic polynomial in $x$ and $y$
\begin{equation}
u=\sum_{0\le i+j\le 2}c_{ij}x^iy^j
\label{eq:sol1}
\end{equation}
is also a solution.

\subsubsection{$2$}
Rotationally invariant solutions are of the form $u=g(s)$ where
$s=x^2+y^2$. The reduced equation is the following ODE for $f(s)$
\begin{equation}
  \label{fullrot}
  (2sg_{ss}+g_s)\left[s(2sg_{ss}+g_s)g_{sss}+(3sg_{ss}+2g_s)g_{ss}\right]^2=0.
\end{equation}
The factorisation of the reduced equation implies that 
\begin{equation}
u=\sqrt{x^2+y^2}, 
\label{eq:sol2}
\end{equation}
is a solution of equation \eqref{eq:infpolylap-2}, which we obtain by solving the linear equation 
$$
2sg_{ss}+g_s=0
$$ and then changing to the original $x,y-$variables. Note that this
solution is also the most general rotationally invariant solution of
Aronsson's equation \eqref{eq:inflap} in two independent
variables \cite{FreireFaleiros:2011}. However, equation \eqref{eq:infpolylap-2} may admit more
solutions of this type that correspond to the equation defined by the
second factor in (\ref{fullrot}), i.e. a third order
nonlinear ODE.

\subsubsection{$3$}
The quantities $u$ and $s=xy^{-1}$ are algebraic invariants of the Lie
group generated by $X_4$. We assume that $u=g(s)$ and we obtain a
reduced differential equation for $g(s)$ which, similarly to the
previous case can be decomposed to a product of two factors. One of these factors is too complicated to include it,
however, the other factor is simpler and defines the differential equation
$$ 
(1+s^2)g_{ss}+2sg_s=0
$$  
from which we can obtain the solution
\begin{equation}
u=\text{arctan}\left(\frac{x}{y}\right),
\label{eq:sol3}
\end{equation}
of equation \eqref{eq:infpolylap-2}. Using the permutation symmetry of the independent variable it follows that $\text{arctan}\left(y/x\right)$ is also a solution. It can be easily verified that any linear combination of these two solutions is again a solution.

\subsubsection{$4$}
In the case of the generator $\gamma X_1+X_5$ we have two subcases
depending on the value of $\gamma$. If $\gamma=0$ the only invariant
solution is the trivial solution $u=0$. If $\gamma=1$ we have two
invariants of the corresponding Lie group, namely $e^{-x}u$ and
$y$. This implies that the most general form of an invariant solution
is $u=e^{x}h(y)$. Substituting the ansatz for $u$ in
\eqref{eq:infpolylap-2} we obtain the following equation for $h(y)$
$$
4hg^2+4h_ygg_y+h_{yy}(g_y)^2=0, \quad g[h]:= h^2+2(h_y)^2+(h_{yy})^2.
$$ 
This equation is difficult to solve and it does not admit any
obvious factorisations as in the previous cases. It is important to note 
at this point that $g[h]=0$ defines a subset of solutions. Obviously, if $h(y)$ is a real function then the only such solution is $h(y)=0$. However, over the complex numbers equation $g[h]=0$ might be tractable and have nontrivial solutions.

\subsubsection{$5$}
In the case of the generator $\alpha X_3+X_5$ the invariants are
$s=x^2+y^2$ and $r=\text{arctan}\left(\frac{x}{y}\right)+\alpha
\ln\left(u\right)$. The most general solution invariant under
the symmetry generated by $\alpha X_3+X_5$ is of the form
$$
u=\exp\left(\frac{1}{\alpha}g(s)-\frac{1}{\alpha}\text{arctan}\left(\frac{x}{y}\right)\right).
$$ The resulting reduced ODE for $g(s)$ is too complicated to handle
or even write down. For specific values of $\alpha$ it might be 
possible to simplify the expressions, due to cancellations or factorisations, and thus find explicit solutions.

\subsubsection{$6$}
The quantities $s=xy^{-1}$ and $r=ux^{-\frac{1}{\alpha}}$ are
invariants under the action of the Lie symmetry generated by $\alpha
X_3+X_5$. In the limit $\alpha\rightarrow\infty$ we reduce to the
generator $X_3$. For $\alpha\neq 0$, the most general invariant
solution of \eqref{eq:infpolylap-2} is of the form
$u=x^{\frac{1}{\alpha}}g(s)$. For a general $\alpha\neq 0$ the reduced ODE it is complicated and we will not present it here. However, it is interesting to note that in the special case $\alpha=1/2$ the reduced equation factorises as follows
$$
E_1[g]^3E_2[g]^2=0
$$
where
$$
E_1[g]:= s^2(1+s^2)^2g_{ss}+2s(1+s^2)(2+s^2)g_s+2 g
$$
and
$$
E_2[g]:= sg_{sss}+6sg_{ss}+6g_s.
$$
Equations $E_1[g]=0$ and $E_2[g]=0$ are both linear and can be solved exactly. Solving the first equation we obtain the following solutions for \eqref{eq:infpolylap-2}
\begin{equation}
u=(x^2+y^2)\left[
c_1\text{cos}\left( \sqrt{2}\text{arctan}\left(\frac{x}{y}\right)\right)+c_2\text{sin}\left( \sqrt{2}\text{arctan}\left(\frac{x}{y}\right)\right)\right]
\label{eq:sol4}
\end{equation}
for $c_i\in\mathbb{R}$. The general solution of equation $E_2[g]=0$ can also be find and implies the solution
$$
u=c_1x^2+c_2xy+c_3y^2
$$ 
with $c_i\in\mathbb{R}$. It would be very interesting to find a method or some criteria which will detect possible values of the parameter in which such factorisations occur. Perhaps such suitable necessary conditions for the parameter $\alpha$ can be obtained using a Painlev\'e type analysis, see Chapter 7 of \cite{ablowitz1991solitons} and \cite{conte2012} for reviews and detailed references.   It is interesting to notice that while the scaling $(x,y,u)\mapsto (e^{\alpha \epsilon}x,e^{\alpha \epsilon}y,e^{\epsilon}u)$ is a symmetry for every $\alpha$, each of the differential monomials of equation \eqref{eq:infpolylap-2} has the same weight, i.e.,
$$
f_x^2u_{xx}\mapsto e^{(5-12 \alpha)\epsilon}f_x^2u_{xx}
$$
and similarly for the other terms. This observation implies that for the special value $\alpha=12/5$ all three terms of equation \eqref{eq:infpolylap-2} are individually invariant under the scaling symmetry. This observation further implies that
$$
\Pi_{\infty}^2x^r\sim x^{5r-12}
$$
and because of symmetry the same will hold for $y^r$. Putting all these together it can be verified that
$$
\Pi_{\infty}^2(ax^r+by^r)\sim a^5x^{5r-12}+b^5y^{5r-12}
$$ from where we obtain, for $r=12\slash 5$, a scaling invariant
solution, also known as \textit{similarity solution}, of equation \eqref{eq:infpolylap-2} if and only if $(a,b)$
satisfies
$$
a^5+b^5=0.
$$
The only real solution is given by $b=-a$ and in this way we recover the solution
\begin{equation}
  \label{eq:sol5}
  u = x^{12\slash 5}-y^{12\slash 5},
\end{equation}
which was first constructed in \cite{KatzourakisPryer:2016}. The same arguments are valid in the case of the general
$\infty-$Polylaplacian in $n$ independent variables. In this case we obtain that
\begin{equation}\label{eq:weeksol-n}
u=c_1x_1^{12\slash 5}+\cdots+c_nx_n^{12\slash 5}
\end{equation}
is a solution if and only if $(c_1,...,c_n)$ lies on the affine variety $V(c_1^5+\cdots+c_n^5)$. This is an invariant solution under the scaling symmetry generated by the vector field
$$
X=\frac{\partial}{\partial x_1}+\cdots+\frac{\partial}{\partial x_n}+\frac{12}{5}\frac{\partial}{\partial u}.
$$
Indeed, it can be written in the following form
$$
I_0=c_1I_1^{12/5}+\cdots+c_{n-1}I_{n-1}^{12/5}+c_n
$$
where 
$$
I_0=ux_n^{-12/5},\quad I_j=x_jx_n^{-1},~j=1,\cdots,n-1
$$
are invariants of the scaling symmetry generated by $X$.
 
\subsubsection{7} For the generator $\gamma X_1+\alpha X_3+X_4$ we have to consider the two subcases $\gamma=0$ and $\gamma=1$. When $\gamma=0$ the generator is $\alpha X_3+X_5$ with $\alpha\in\mathbb{R}\backslash\lbrace 0\rbrace$ and we can verify that $u$ and
$$
s=\text{arctan}\left(\frac{x}{y}\right)-\frac{\alpha}{2}\ln\left(x^2+y^2\right)
$$
are invariants under the action of the corresponding Lie group. The ansatz $u=g(s)$ leads to an ODE that is too complicated to include it here. Nevertheless, the reduced ODE admits a factorisation where one of the factors is given by
$$
E_1[g]:=(1+\alpha^2)g_{ss}+\alpha g_s.
$$ 
The linear ODE $E_1[g]=0$ can be integrated for every $\alpha$ and gives the following solution of the $\infty$-Polylaplacian
\begin{equation}
u=\frac{(x^2+y^2)^{\frac{\alpha^2}{2(1+\alpha^2)}}}{\exp\left(\frac{\alpha}{1+\alpha^2}\text{arctan}\left(\frac{x}{y}\right)\right)}
\label{eq:sol6}
\end{equation}
Similarly, when $\gamma=1$ the invariants are $u$ and
$$
z=-4\text{arctan}\left(\frac{\alpha x+y}{1+x-\alpha y}\right)+2\alpha\ln\left[
\alpha^2+2\alpha^2 x-2\alpha^3y+\alpha^2(1+\alpha^2)(x^2+y^2)
\right]
$$
and the reduced ODE for $h(z)$ contains the following linear factor
$$
E_2[h]:= 4(1+\alpha^2)h_{zz}-\alpha h_z.
$$
Solving the linear ODE $E_2[h]=0$ for all $\alpha$ we finally obtain the following solution
\begin{equation}
u=\frac{\left[
\alpha^2+2\alpha^2 x-2\alpha^3y+\alpha^2(1+\alpha^2)(x^2+y^2)
\right]^{\frac{\alpha^2}{2(1+\alpha^2)}}}{\exp\left(
\frac{\alpha}{1+\alpha^2}\text{arctan}\left(\frac{\alpha x+y}{1+x-\alpha y}\right)
\right)}
\label{eq:sol7}
\end{equation}
of the $\infty$-Polylaplacian. It is interesting to notice that solution \eqref{eq:sol6} can be seen as the dominant part of solution \eqref{eq:sol7} as $\alpha\rightarrow\infty$.

Unfortunately the reductions that correspond to the generator $A8$ are
too complicated to handle. We now focus on the reductions of the
reduced $\infty-$Polylaplacian \eqref{eq:reducedinf} that give
additional information. Solutions of equation \eqref{eq:reducedinf}
that are invariant under translation in the $x$-direction are of the
form $u=g(y)$ where $f$ satisfies
$$
0=g_{yy}^2-1=(g_{yy}+1)(g_{yy}-1).
$$
The solutions of these equations are just quadratic polynomials in $y$ and thus add nothing new. Solutions invariant under rotations are of the form $u=g(s)$ where $s=x^2+y^2$ and $g(s)$ satisfies the following ODE
$$
16 s^2g_{ss}^2+16sg_sg_{ss}+8g_s^2-1=0.
$$
The general solution of this ODE is not known, nevertheless a simple polynomial ansatz can lead to the special solution $s/2\sqrt{2}$. The corresponding solution of the \eqref{eq:reducedinf} and hence of \eqref{eq:infpolylap-2} is contained in the family of polynomial solutions. Finally, the ansatz $u=x^2 g(s)$ where $s=xy^{-1}$ leads to solutions that are invariant under the Lie symmetry generated by $Y_4$. In this case the reduced ODE for $g$ is given by
$$
s^4(1+s^2)^2g_{ss}^2+4s^2(g+s(2+3s^2+s^4)g_s)g_{ss}+2s^2(8+9s^2+2s^4)g_s^2+16sgg_s+4g^2=1.
$$
As before a Laurent polynomial ansatz leads to the special solutions $
(\sqrt{2}s)^{-1}$ and $(2s^2)^{-1}$.
The corresponding solutions of the $\infty-$Polylaplacian are contained in the polynomial family. Due to the complexity of the expressions we didn't manage to obtain something meaningful in the final case $(B4)$.

\section{Conclusions and discussion}

In this paper we studied the $\infty$-Polylaplacian equation
\eqref{eq:infpolylap-2} and the reduced $\infty$-Polylaplacian
equation \eqref{eq:reducedinf} in two dimensions ($n=2$) from an
algebraic point of view. The latter can be seen as a second order
analogue of the Eikonal equation. For both equations we found the
complete group of Lie point symmetries and we classified all the
non-equivalent, under the adjoint action, one dimensional Lie
subalgebras {of $\widetilde{\mathfrak{g}}$ and
  $\widetilde{\mathfrak{h}}$ that correspond to translations,
  rotations and scalings}. For each generator in our list we
constructed canonical invariant coordinates and used them to perform
the corresponding symmetry reduction. We studied the obtained reduced
ODEs and constructed many new self-similar special solutions
\eqref{eq:sol1}, \eqref{eq:sol2}, \eqref{eq:sol3}, \eqref{eq:sol4},
\eqref{eq:sol5}, \eqref{eq:sol6}, \eqref{eq:sol7}. The family of
functions \eqref{eq:weeksol-n}, parametrised by an affine variety, is
a solution of \eqref{eq:infpolylap-n} for every $n$. {The
  complete classification of one dimensional subalgebras of the full
  symmetry algebras $\mathfrak{g}$ and $\mathfrak{h}$ together with a
  more in depth analysis of all of their invariant solutions is still
  an open problem and is left for future work.} It is also interesting to
investigate the structure of the solutions for $n\ge 3$. As a first
step towards this direction we presented a conjecture on the full
group of Lie point symmetries of the $\infty$-Polylaplacian and its
reduced version in $n$-dimensions.

We believe that for this type of strongly nonlinear PDE that arise
in calculus of variations in $L^{\infty}$ deep intuition can be
gained by studying the structure of their Lie symmetries. There are
also many related open problems. For example, currently, to the best
of the author's knowledge, a Noether-like theorem for these problems
is not known. A topic of ongoing work is to investigate whether
Noether's classical theorem applied in variational problems in $L^p$
survives the limit $p\rightarrow\infty$.

\bibliographystyle{alpha}
\bibliography{tristansbib,tristanswritings}

\begin{thebibliography}{BDM89}

\bibitem[AC91]{ablowitz1991solitons}
Mark~J Ablowitz and Peter~A Clarkson.
\newblock {\em Solitons, nonlinear evolution equations and inverse scattering},
  volume 149.
\newblock Cambridge university press, 1991.

\bibitem[Aro65]{Aronsson:1965}
G.~Aronsson.
\newblock Minimization problems for the functional \newline {${\rm
  sup}_{x}\,F(x,\,f(x),\,f^{\prime} (x))$}.
\newblock {\em Ark. Mat.}, 6:33--53 (1965), 1965.

\bibitem[Aya18]{Ayanbayev:2018}
Birzhan Ayanbayev.
\newblock Explicit $\infty$-harmonic functions in high dimensions.
\newblock {\em Journal of Elliptic and Parabolic Equations}, Jun 2018.

\bibitem[BA08]{Bluman2008}
G.~Bluman and S.~Anco.
\newblock {\em Symmetry and integration methods for differential equations},
  volume 154.
\newblock Springer Science \& Business Media, 2008.

\bibitem[Bar99]{barron1999}
N.~Barron.
\newblock Viscosity solutions and analysis in $l^{\infty}$.
\newblock {\em Nonlinear Analysis, Differential Equations and Control, Nato
  Science Series C}, 528, 1999.

\bibitem[BC69]{Bluman1969}
G.~Bluman and J.~Cole.
\newblock The general similarity solution of the heat equation.
\newblock {\em Journal of Mathematics and Mechanics}, 18(11):1025--1042, 1969.

\bibitem[BDM89]{BhattacharyaDiBenedettoManfredi:1989}
T.~Bhattacharya, E.~DiBenedetto, and J.~Manfredi.
\newblock Limits as $p\to\infty$ of $\delta_p u_p = f$ and related extremal
  problems.
\newblock {\em Rend. Sem. Mat. Univ. Politec. Torino}, 47:15--68, 1989.

\bibitem[BEJ08]{BarronEvansJensen:2008}
E.~N. Barron, L.~C. Evans, and R.~Jensen.
\newblock The infinity {L}aplacian, {A}ronsson's equation and their
  generalizations.
\newblock {\em Trans. Amer. Math. Soc.}, 360(1):77--101, 2008.

\bibitem[BGI89]{Baikov1989}
V.~A. Baikov, R.~K. Gazizov, and N.~K. Ibragimov.
\newblock Approximate symmetries.
\newblock {\em Sbornik: Mathematics}, 64:427--441, 1989.

\bibitem[CLO92]{Cox1992ideals}
D.~Cox, J.~Little, and D.~O'Shea.
\newblock {\em Ideals, varieties, algorithms: An introduction to computational
  algebraic geometry and commutative algebra, UTM}.
\newblock Springer-Verlag, New York, 1992.

\bibitem[CM94]{Clarkson1994}
P.~A. Clarkson and E.~L. Mansfield.
\newblock Algorithms for the nonclassical method of symmetry reductions.
\newblock {\em SIAM Journal on Applied Mathematics}, 54(6):1693--1719, 1994.

\bibitem[Con12]{conte2012}
Robert Conte.
\newblock {\em The Painlev{\'e} property: one century later}.
\newblock Springer Science \& Business Media, 2012.

\bibitem[DT04]{Dimas2004sym}
S.~Dimas and D.~Tsoubelis.
\newblock Sym: A new symmetry-finding package for mathematica.
\newblock {\em Proceedings of the 10th international conference in modern group
  analysis}, pages 64--70, 2004.

\bibitem[DT06]{Dimas2006}
S.~Dimas and D.~Tsoubelis.
\newblock A new mathematica-based program for solving overdetermined systems of
  pdes.
\newblock {\em 8th International Mathematica Symposium}, 2006.

\bibitem[ETT15]{ElmoatazToutainTenbrinck:2015}
Abderrahim Elmoataz, Matthieu Toutain, and Daniel Tenbrinck.
\newblock On the $p$-{L}aplacian and $\infty$-{L}aplacian on graphs with
  applications in image and data processing.
\newblock {\em SIAM Journal on Imaging Sciences}, 8(4):2412--2451, 2015.

\bibitem[EY05]{evans2005}
L~C Evans and Y~Yu.
\newblock Various properties of solutions of the infinity-laplacian equation.
\newblock {\em Comm Partial Differential Equations}, 30:1401--1428, 2005.

\bibitem[FF11]{FreireFaleiros:2011}
Igor~Leite Freire and Antonio~C{\^a}ndido Faleiros.
\newblock Lie point symmetries and some group invariant solutions of the
  quasilinear equation involving the infinity laplacian.
\newblock {\em Nonlinear Analysis: Theory, Methods \& Applications},
  74(11):3478--3486, 2011.

\bibitem[For90]{fordy}
Allan~P Fordy.
\newblock {\em Soliton theory: a survey of results}.
\newblock Manchester University Press, 1990.

\bibitem[GM10]{GyulovMorosanu:2010}
Tihomir Gyulov and Gheorghe Moro{\c{s}}anu.
\newblock On a class of boundary value problems involving the $p$-{B}iharmonic
  operator.
\newblock {\em Journal of Mathematical Analysis and Applications},
  367(1):43--57, 2010.

\bibitem[Her97]{hereman1997}
W.~Hereman.
\newblock Review of symbolic software for lie symmetry analysis.
\newblock {\em Mathematical and Computer Modelling}, 25:115--132, 1997.

\bibitem[Hyd00]{Hydon2000}
Peter~E Hydon.
\newblock {\em Symmetry methods for differential equations: a beginner's
  guide}, volume~22.
\newblock Cambridge University Press, 2000.

\bibitem[Ibr93]{Ibragimov1993crc}
Nail~H Ibragimov.
\newblock {\em CRC Handbook of Lie group analysis of differential equations},
  volume 1-3.
\newblock CRC press, 1993.

\bibitem[Igb12]{Igbida:2012}
Noureddine Igbida.
\newblock A partial integrodifferential equation in granular matter and its
  connection with a stochastic model.
\newblock {\em SIAM Journal on Mathematical Analysis}, 44(3):1950--1975, 2012.

\bibitem[Jen93]{Jensen:1993}
Robert Jensen.
\newblock Uniqueness of {L}ipschitz extensions: minimizing the sup norm of the
  gradient.
\newblock {\em Arch. Rational Mech. Anal.}, 123(1):51--74, 1993.

\bibitem[Kat15]{Katzourakis:2015}
Nikos Katzourakis.
\newblock {\em An introduction to viscosity solutions for fully nonlinear {PDE}
  with applications to calculus of variations in {$L^\infty$}}.
\newblock Springer Briefs in Mathematics. Springer, Cham, 2015.

\bibitem[KP16]{KatzourakisPryer:2015}
Nikos Katzourakis and Tristan Pryer.
\newblock On the numerical approximation of $\infty$-harmonic mappings.
\newblock {\em Nonlinear Differential Equations and Applications NoDEA},
  23(6):61, 2016.

\bibitem[KP17]{KatzourakisPryer:2017}
Nikos Katzourakis and Tristan Pryer.
\newblock On the numerical approximation of $p$-biharmonic and
  $\infty$-biharmonic functions.
\newblock {\em arXiv:1701.07415}, 2017.

\bibitem[KP18]{KatzourakisPryer:2016}
Nikos Katzourakis and Tristan Pryer.
\newblock Second order ${L}^\infty$ variational problems and the
  $\infty$-polylaplacian.
\newblock {\em Advances in Calculus of Variations}, 2018.

\bibitem[LM90]{LazerMcKenna:1990}
AC~Lazer and PJ~McKenna.
\newblock Large-amplitude periodic oscillations in suspension bridges: some new
  connections with nonlinear analysis.
\newblock {\em Siam Review}, 32(4):537--578, 1990.

\bibitem[Oli10]{Oliveri2010}
Francesco Oliveri.
\newblock Lie symmetries of differential equations: classical results and
  recent contributions.
\newblock {\em Symmetry}, 2(2):658--706, 2010.

\bibitem[Olv93]{Olver:1993}
Peter~J. Olver.
\newblock {\em Applications of {L}ie groups to differential equations}, volume
  107 of {\em Graduate Texts in Mathematics}.
\newblock Springer-Verlag, New York, second edition, 1993.

\bibitem[Ovs82]{Ovsiannikov1982}
LV~Ovsiannikov.
\newblock Group analysis of differential equations.
\newblock {\em Academic Press, New York}, 1982.

\bibitem[Pry17]{Pryer:2015}
Tristan Pryer.
\newblock On the finite element approximation of infinity-harmonic functions.
\newblock {\em To appear in {P}roceedings {A} of the {R}oyal {S}ociety of
  {E}dinburgh.}, 2017.

\bibitem[PWZ75]{PateraWinternitzZassenhaus:1975}
Ji{\v{r}}{\'\i} Patera, Pavel Winternitz, and Hans Zassenhaus.
\newblock Continuous subgroups of the fundamental groups of physics. i. general
  method and the poincar{\'e} group.
\newblock {\em Journal of Mathematical Physics}, 16(8):1597--1614, 1975.

\bibitem[Rei90]{Reid1990}
GJ~Reid.
\newblock A triangularization algorithm which determines the lie symmetry
  algebra of any system of pdes.
\newblock {\em Journal of Physics A: Mathematical and General}, 23(17):L853,
  1990.

\bibitem[Sch07]{Schwarz2007}
Fritz Schwarz.
\newblock {\em Algorithmic Lie theory for solving ordinary differential
  equations}.
\newblock CRC Press, 2007.

\bibitem[Ste89]{Stephani1989}
Hans Stephani.
\newblock {\em Differential equations: their solution using symmetries}.
\newblock Cambridge University Press, 1989.

\bibitem[{\v{S}}W14]{SnoblWinternitz:2014}
Libor {\v{S}}nobl and Pavel Winternitz.
\newblock {\em Classification and identification of Lie algebras}, volume~33.
\newblock American Mathematical Soc., 2014.

\bibitem[Win90]{Winternitz:1990}
P~Winternitz.
\newblock Group theory and exact solutions of partially integrable differential
  systems.
\newblock In {\em Partially Intergrable Evolution Equations in Physics}, pages
  515--567. Springer, 1990.

\bibitem[Win93]{winternitz1993lie}
Pavel Winternitz.
\newblock Lie groups and solutions of nonlinear partial differential equations.
\newblock {\em Nato Asi Series C Mathematical And Physical Sciences}, 409,
  1993.

\bibitem[Yag88]{Yaglom1988}
M.~Yaglom.
\newblock Felix klein and sophus lie.
\newblock {\em Evolution of the Idea of Symmetry in the Nineteenth Century,
  Birkh{\"a}user, Boston etc}, 1988.

\end{thebibliography}

\end{document}